\documentclass[aps,prd,groupedaddress,showpacs,showkeys,nofootinbib]{revtex4}%
\usepackage{epsfig,amssymb,amsmath}
\usepackage[english]{babel}%
\usepackage{amsmath}%
\usepackage{amsfonts}%
\usepackage{amssymb}%
\usepackage{graphicx}
\newcommand{\beq}{\begin{equation}}
\newcommand{\eeq}{\end{equation}}
\newcommand{\bea}{\begin{eqnarray}}
\newcommand{\eea}{\end{eqnarray}}
\newcommand{\beaa}{\begin{eqnarray}}
\newcommand{\eeaa}{\end{eqnarray}}
\newcommand{\ba}{\begin{array}}
\newcommand{\ea}{\end{array}}
\newcommand{\bit}{\begin{itemize}}
\newcommand{\eit}{\end{itemize}}
\newcommand{\ben}{\begin{enumerate}}
\newcommand{\een}{\end{enumerate}}

\def\1{{_{1}}}
\def\2{{_{2}}}

\begin{document}
\title{Dark energy and dust matter phases from an exact $f(R)$-cosmology model}
\author{S.Capozziello${}^{\sharp}$, P. Martin-Moruno${}^{\flat}$, C. Rubano${}%
^{\sharp}$}
\affiliation{}
\affiliation{${}^{\sharp}$ Dipartimento di Scienze Fisiche, Universit\`{a} di Napoli
"Federico II" and INFN Sez. di Napoli, Compl. Univ. Monte S. Angelo, Ed.N, Via
Cinthia, I-80126 Napoli, Italy,}
\affiliation{${}^{\flat}$ Colina de los Chopos, Instituto de Matematicas y Fisica
Fundamental, Consejo Superior de Investigaciones Cientificas, Serrano 121,
28006 Madrid, Spain.}
\date{\today}

\begin{abstract}
We show that dust matter-dark energy combined phases can be
achieved by the exact solution derived from a power law $f(R)$
cosmological model. This example answers the query by which a
dust-dominated decelerated phase, before dark-energy accelerated
phase, is needed in order to form large scale structures.

\end{abstract}
\maketitle

%\vspace{2mm}

%\maketitle

Very recently, alternative theories of gravity are playing an interesting role
to describe the today observed Universe. Although being the best fit to a wide
range of data \cite{LambdaTest}, the $\Lambda$CDM model is affected by strong
theoretical shortcomings \cite{LambdaRev} that have motivated the search for
alternative models \cite{PR03,copeland}.

Dark energy models mainly rely on the implicit assumption that Einstein's GR
is the correct theory of gravity indeed. Nevertheless, its validity on large
astrophysical and cosmological scales has never been tested but only
\textit{assumed} \cite{will}, and it is therefore conceivable that both cosmic
speed up and missing matter are nothing else but signals of a breakdown of GR.
In this sense, GR could fail in giving self-consistent pictures both at
ultraviolet scales (early universe) and at infrared scales (late universe).

Following this line of thinking, the "minimal" choice could be to take into
account generic functions $f(R)$ of the Ricci scalar $R$. The task for this
extended theories should be to match the data under the "economic" requirement
that no exotic dark ingredients have to be added, unless these are going to be
found by means of fundamental experiments \cite{kleinert,noi}. This is the
underlying philosophy of what are referred to as \textit{$f(R)$-gravity} (see
\cite{copeland,odirev,GRGrev} and references therein).

Although higher order gravity theories have received much attention in
cosmology, since they are naturally able to give rise to the accelerating
expansion (both in the late and in the early universe \cite{noi}), it is
possible to demonstrate that $f(R)$ theories can also play a major role at
astrophysical scales. In fact, modifying the gravity Lagrangian affects the
gravitational potential in the low energy limit. Provided that the modified
potential reduces to the Newtonian one on the Solar System scale, this
implication could represent an intriguing opportunity rather than a
shortcoming for $f(R)$ theories. In fact, a corrected gravitational potential
could offer the possibility to fit galaxy rotation curves without the need of
huge amounts of dark matter
\cite{noipla,mond,jcap,mnras,sobouti,salucci,mendoza}. In addition, it is
possible to work out a formal analogy between the corrections to the Newtonian
potential and the usually adopted galaxy halo models which allow to reproduce
dynamics and observations \textit{without} dark matter \cite{jcap}.

However, extending the gravitational Lagrangian could give rise to several
problems. These theories could have instabilities \cite{instabilities-f(R)},
ghost\,-\,like behaviors \cite{ghost-f(R)}, and they have to be matched with
the low energy limit experiments which quite fairly test GR.

In summary, it seems that the paradigm to adopt $f(R)$-gravity leads to
interesting results at cosmological, galactic and Solar System scales but, up
to now, no definite physical criterion has been found to \textit{select} the
final $f(R)$ theory (or class of theories) capable of matching the data at all
scales. Interesting results have been achieved in this line of thinking
\cite{mimicking,Hu,Star,Odintsov1} but the approaches are all phenomenological
and are not based on some fundamental principle as the conservation or the
invariance of some quantity or some intrinsic symmetry of the theory.

In some sense, the situation is similar to that of dark matter: we know very
well its effect at large astrophysical scales but no final evidence of its
existence has been found, up to now, at fundamental level. In the case of
$f(R)$-gravity, we know that the paradigm is working: in principle, the
missing matter and accelerated cosmic behavior can be addressed taking into
account gravity (in some extended version), baryons and radiation but we do
not know a specific criterion to select the final, comprehensive theory.

In this letter, we want to show that a general exact solution, coming from the
request of the existence of a Noether symmetry for $f(R)$ cosmological models,
matches the two main important requirements that a cosmological solution
should achieve to agree with data: a transient Friedmann dust-like phase,
needed for structure formation, and an asymptotic accelerated behavior. Far to
be the final model to explain the cosmic speed up, the presence of the Noether
symmetry could be a physically motivated approach to select viable
cosmological models.

The general features of the theory are the following. Let
\begin{equation}
\mathcal{A}=\int d^{4} x\,\sqrt{-g}\,f(R)+\mathcal{A}_{m}\, ,
\end{equation}
be the gravitational action where $f(R)$ is a generic function of the Ricci
scalar $R$. GR is recovered in the particular case $f(R)=-R/16\pi G$, and
$\mathcal{A}_{m}$ is the action for a perfect fluid minimally coupled with gravity

In the metric formalism, this action leads to 4th order differential
equations
\begin{equation}
\label{field}f_{R}\,R_{\mu\nu}-\tfrac12\,f\,g_{\mu\nu}-f_{R;\mu\nu}+g_{\mu\nu
}\,\Box f_{R}=-\tfrac12\,T^{m}_{\mu\nu}\, ,
\end{equation}
where a subscript $R$ denotes differentiation with respect to $R$ and
$T^{m}_{\mu\nu}$ is the matter fluid stress-energy tensor.

In order to derive the cosmological equations in a Friedman-Robertson-Walker
(FRW) metric, one can define a canonical Lagrangian $\mathcal{L}%
=\mathcal{L}(a,\dot{a}, R, \dot{R})$, where $\mathcal{Q}=\{a,R\}$ is the
configuration space and $\mathcal{TQ}=\{a,\dot{a}, R, \dot{R}\}$ is the
related tangent bundle on which $\mathcal{L}$ is defined. The variable $a(t)$
and $R(t)$ are the scale factor and the Ricci scalar in the FRW metric,
respectively. One can use the method of the Lagrange multipliers to set $R$ as
a constraint of the dynamics. Selecting the suitable Lagrange multiplier and
integrating by parts, the Lagrangian $\mathcal{L}$ becomes canonical. In our
case, we have%

\begin{equation}
\mathcal{A}=2\pi^{2}\int dt\,a^{3} \left\{ f(R)-\lambda\left[ R+6\left(
\frac{\ddot a}a+\frac{\dot a^{2}}{a^{2}}+\frac k{a^{2}}\right) \right]
\right\}  ,
\end{equation}
It is straightforward to show that, for $f(R)=-R/16\pi G$, one obtains the
usual Friedman equations.

The variation with respect to $R$ of the Lagrange multiplier gives
$\lambda=f_{R}$. Therefore, integrating by parts, the point-like FRW
Lagrangian is
\begin{equation}
\mathcal{L}= a^{3}\,(f-f_{R}\,R)+6\,a^{2}\,f_{RR}\,\dot R\,\dot a
+6\,f_{R}\,a\,\dot a^{2}-6k\,f_{R}\,a\, ,\label{eqz0}%
\end{equation}
which is a canonical function of two coupled fields, $R$ and $a$, both
depending on time $t$. The total energy $E_{\mathcal{L}}$, corresponding to
the $\{0,0\}$-Einstein equation, is
\begin{equation}
\label{energy}E_{\mathcal{L}}=6\,f_{RR}\,a^{2}\,\dot a\,\dot R+ 6\,f_{R}%
\,a\,\dot a^{2}- a^{3}\,(f-f_{R}\,R) +6k\,f_{R}\,a=D\, .
\end{equation}
where $D$ represents the standard amount of dust fluid as, for example,
measured today. The equations of motion for $a$ and $R$ are respectively
\begin{align}
f_{RR}\left[ R+6\,H^{2}+6\,\frac{\ddot a}a+6\,\frac{k}{a^{2}}\right]  &
=0\label{frd1}\\
6\,f_{RRR}\,\dot R^{2}+ 6\,f_{RR}\,\ddot R+ 6\,f_{R}\,H^{2}+12\,f_{R}%
\,\frac{\ddot a}a & = 3\,(f-f_{R}\,R)-12\,f_{RR}\,H\,\dot R -6\,f_{R}%
\,\frac{k}{a^{2}}\, ,\label{frd2}%
\end{align}
where $H\equiv\dot a/a$ is the Hubble parameter. Considering $R$ and $a$ as
independent variables, we have, for consistency, that $R$ coincides with the
definition of the Ricci scalar in the FRW metric.

The form of the function $f(R)$ and the solution of the system (\ref{energy}),
(\ref{frd1}) and (\ref{frd2}) can be achieved by asking for the existence of
Noether symmetries. On the other hand, the existence of the Noether symmetries
guarantees the reduction of dynamics and the eventual solvability of the
system \cite{cimento,leandros,lambiase}. Here, we want to seek for viable
$f(R)$ cosmological models.

We shall focus our attention on the fact that we need a
cosmological solution of the field equations which exhibits not
only an accelerated phase in recent universe, but also a
decelerated period, which lasts for a long time, sufficient to
allow the formation of structures. This issue has recently been
argument of debate since the validity of $f(R)$ cosmology, which
claims to avoid unknown ingredient as dark energy, strictly lies
on this possibility \cite{CNOT}. Several works on $f(R)$-gravity
have been devoted to the acceleration and the reconstruction of
the models starting from data \cite{noi}. Numerical treatment is
almost obliged and some educated, although arbitrary, guess on the
functional form is often necessary. On the other hand,
$f(R)$-cosmology should give rise to standard Friedmann
dust-dominated phase, which is necessary for the structure
formation mechanism, widely accepted and properly working. A first
answer to this issue was given by means of a numerical
reconstruction of the $f(R)$ function \cite{mimicking}. Here, we
want to present a general exact solution of the equations,
obtained by means of the so called \ \textquotedblleft Noether
Symmetry Approach\textquotedblright. A summary of the method can
be found in \cite{cimento,leandros,lambiase}.

We ask now for the existence of a vector field
\begin{equation}
X=\alpha\frac{\partial}{\partial a}+\beta\frac{\partial}{\partial R}%
+\dot{\alpha}\frac{\partial}{\partial\dot{a}}+\dot{\beta}\frac{\partial
}{\partial\dot{R}},
\end{equation}
such that the Lie derivative of the Lagrangian is zero, i.e. $\mathcal{L}$ is
conserved and $X$ is a Noether symmetry. It is then possible to find%
\begin{equation}
\label{sym}\alpha=1/a\quad;\quad\beta=-2R/a^{2}\quad;\quad f(R)=-\left\vert
R\right\vert ^{3/2}.
\end{equation}

The absolute value is needed, because (with our conventions) we have $R<0$.
\ Once the symmetry is found, we have an additional constant of the motion,
and it is then easy to find a change of variables $\{a,R\}\rightarrow\{u,v\}$,
such that one of the variables is cyclic. We have in fact%
\begin{equation}
u=a^{2}\left\vert R\right\vert \quad;\quad v=a^{2}/2
\end{equation}
and the new Lagrangian is%
\begin{equation}
\mathcal{L}^{\prime}=\frac{u^{3/2}}{2}+\frac{9}{2}\frac{\dot{u}\dot{v}}%
{\sqrt{u}}-9k\sqrt{u}.\label{int_5}%
\end{equation}

The Noether charge is then $\Sigma_{1}=\dot{u}/\sqrt{u}$, leading to immediate
integration for $u$. Introducing the solution into $E_{\mathcal{L}}=D$, and
solving for $v$ we obtain
\begin{align}
u  &  =\frac{1}{4}\left(  \Sigma_{1}t+\Sigma_{0}\right) ^{2} \quad
\label{int_6}\\
\quad v  &  =\frac{\Sigma_{1}^{2}}{288}t^{4}+\frac{\Sigma_{1}\Sigma_{0}}%
{72}t^{3}+\left(  \frac{\Sigma_{0}^{2}}{48}-\frac{k}{2}\right)  t^{2}+\left(
\frac{\Sigma_{0}^{3}}{72\Sigma_{1}}-k\frac{\Sigma_{0}}{\Sigma_{1}}+\frac
{2D}{9\Sigma_{1}}\right)  t+v_{0}.
\end{align}

The parameters $\Sigma_{0}$, $\Sigma_{1}$, $D$, and $v_{0}$ are the
integration constants of the equations. They are four since this is a general
solution of a fourth order problem.

Coming back to $a(t)$, and setting, for the sake of simplicity $a(0)=0$, i.e.
$v_{0}=0$, we get%
\begin{equation}
\label{solution}a=\sqrt{a_{4}t^{4}+a_{3}t^{3}+a_{2}t^{2}+a_{1}t},
\end{equation}
with%
\[
a_{4}=\frac{\Sigma_{1}^{2}}{144}\quad;\quad a_{3}=\frac{\Sigma_{1}\Sigma_{0}%
}{36}\quad;\quad a_{2}=\frac{\Sigma_{0}^{2}}{24}-k\quad;\quad a_{1}%
=\frac{\Sigma_{0}^{3}}{36\Sigma_{1}}-2k\frac{\Sigma_{0}}{\Sigma_{1}}+\frac
{4D}{9\Sigma_{1}}.
\]

%\section{Fixing the parameters.}

We see that this solution is $a\propto t^{2}$ for large $t$, and $a\propto
t^{1/2}$, for small $t$. There is thus room for a smooth transition, passing
through a period during which the solution approximates reasonably well a
Friedmann dust-transient like $a_{f}\propto t^{2/3}$. In order to see this, we
have to consider suitable values of the integration constants $a_{i}$. All
computations and results are simplified if we fix the time unit, by setting
the current time $t_{0}=1$. This will not affect the results but the value of
$H_{0}$ has to be recast with respect to physical units. We assume also
$H_{0}=1$ for simplicity. A unitary value for $a_{0}$ can be also set, if no
restriction on the value of $k$ is imposed. Finally, we consider a value of
the deceleration parameter $q_{0}=-0.4$, which could describe a reasonable
current acceleration. These considerations yield a model depending only on one
parameter. Taking $a_{4}=0.106$, the scale factor turn out to be expressed as
\begin{equation}
a=\sqrt{\frac{t}{5}[2+0.53(t-1)^{3}+t+2t^{2}]}.\label{sec2_1}%
\end{equation}

\begin{figure}
[ptb]
\begin{center}
\includegraphics[
height=2.2433in,
width=3.3797in
]%
{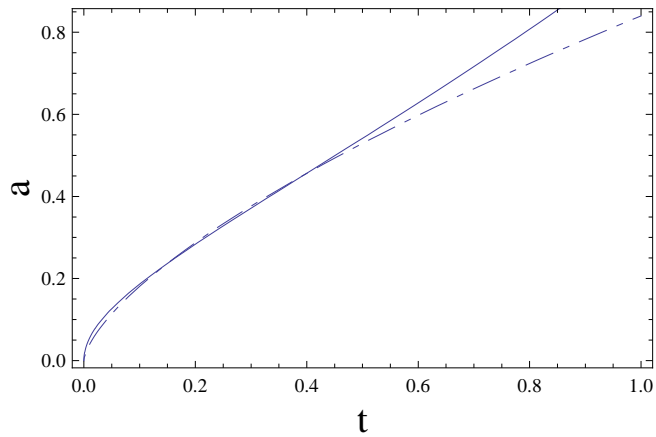}
 \caption{Scale factor versus time in standard model
(dashed) and our model\label{figura1}
(continuous).}%
\end{center}
\end{figure}

Comparing this solution with $a_{f}=a_{0f}t^{2/3}$ and noting that
$a_{0f}$ must be less than $a_{0}$, we obtain the very good
coincidence of Fig.\ref{figura1}. The difference is close to 3\%
in the interval $2\leq
z\leq4$, enough for a phase dominated by galaxies.%
%TCIMACRO{\FRAME{ftbpFU}{3.3797in}{2.2433in}{0pt}{\Qcb{Scale factor versus time
%in standard model (dashed) and our model (continous).}}{}{figura1.eps}%
%{\special{ language "Scientific Word";  type "GRAPHIC";
%maintain-aspect-ratio TRUE;  display "USEDEF";  valid_file "F";
%width 3.3797in;  height 2.2433in;  depth 0pt;  original-width 3.333in;
%original-height 2.2035in;  cropleft "0";  croptop "1";  cropright "1";
%cropbottom "0";  filename 'figura1.eps';file-properties "XNPEU";}}}%
%BeginExpansion
%EndExpansion

It is interesting to come back to the original parameters, in particular for
what is concerning the spatial curvature. We have $k\simeq-0.49$, which yields
$\Omega_{k,0}=kG_{eff}/(3H_{0}^{2}a_{0}^{2})\simeq-0.02$, with $G_{eff}%
=1/[2f^{\prime}(R)]$. Therefore, this model describes a spatially open
universe instead of a spatially flat $k=0$. Indeed, what is physically
relevant is not the value of $k$ , which is connected with the normalization
of $a$, but the dimensionless parameter $\Omega_{k}$. Moreover, the alleged
statement $\Omega_{k}\simeq0$, is obtained from the spectrum of the CMBR
radiation and strongly depends on the standard $\Lambda$CDM model. Another
relevant parameter is the matter content. With our choice of the parameters we
get $D\simeq0.88$, this value implies $\Omega_{m,0}\simeq0.042$, which is very
close to the expected content of baryonic matter in the Universe. One could
consider an observer living within a universe described by our model. If this
observer is unaware of the fact that the function $f(R)$ in the Lagrangian is
$f(R)=-\left\vert R\right\vert ^{3/2}$ and not $f(R)=R$, he would perform all
calculations taking into account $G_{N}$ (and not with $G_{eff}$), obtaining
$\Omega_{m,0}^{\prime}\sim0.29$. This value is the expected one for all the
matter content in the Universe, included the dark matter. Therefore, in this
framework, it seems that taking into account dark matter could be nothing else
but an assumption due to the ignorance of the physical theory behind the
cosmological model.

It can also be noted that $\Omega_{m,0}$ has nearly the same value of
$-\Omega_{k,0}$. Since we have $\Omega_{m,0}+\Omega_{k,0}+\Omega_{R,0}=1$, the
current dynamic of this universe results almost totally driven by the
curvature, being $\Omega_{R,0}\simeq0.98$.%

%TCIMACRO{\FRAME{ftbpFU}{3.3797in}{2.2278in}{0pt}{\Qcb{Percentage difference
%$\delta a$ of the two scale factors, for a range in time corresponding to
%$z=2\div4$. It is less than 3\%.}}{}{figura2.eps}%
%{\special{ language "Scientific Word";  type "GRAPHIC";
%maintain-aspect-ratio TRUE;  display "USEDEF";  valid_file "F";
%width 3.3797in;  height 2.2278in;  depth 0pt;  original-width 3.333in;
%original-height 2.188in;  cropleft "0";  croptop "1";  cropright "1";
%cropbottom "0";  filename 'figura2.eps';file-properties "XNPEU";}}}%
%BeginExpansion
\begin{figure}
[ptb]
\begin{center}
\includegraphics[
height=2.2278in,
width=3.3797in
]%
{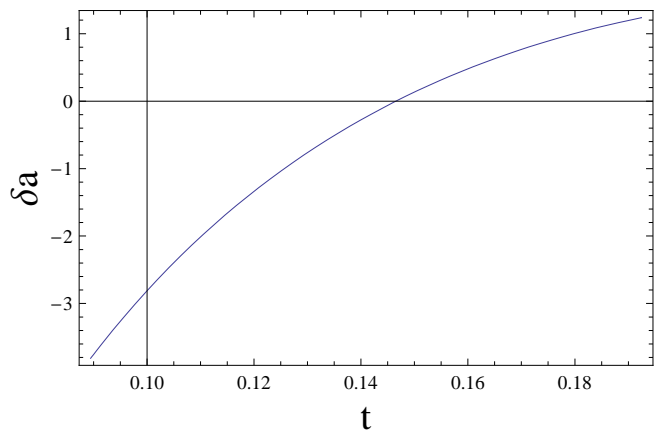}%
\caption{Percentage difference $\delta a$ of the two scale
factors, for a range in time corresponding to $z=2\div4$. It is
less than 3\%.}\label{figura2}%
\end{center}
\end{figure}
%EndExpansion

\begin{figure}
[ptb]
\begin{center}
\includegraphics[
height=2.2001in,
width=3.3797in
]%
{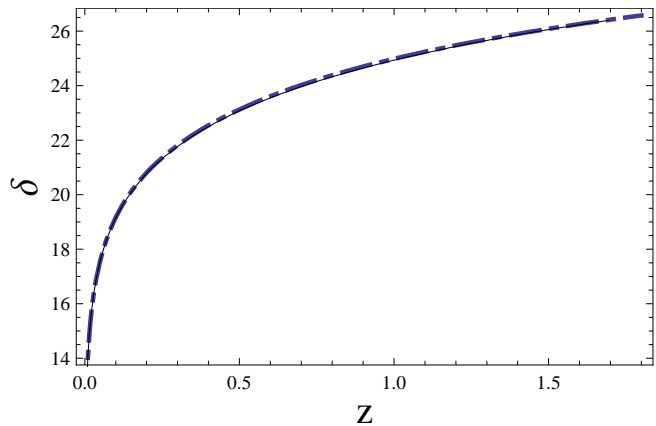}%
\caption{Comparison of the distance modulus $\delta$. Our model
(continous) and $\Lambda CDM$ (dashed). The agreement is almost
perfect.}\label{figura3}%
\end{center}
\end{figure}

In order to check our model in another way, we consider the
distance modulus given by the SNIa and we compare our solution
with the standard $\Lambda$CDM model, as we know that it fits data
very well. Taking as reference the standard solution for
$\Lambda$CDM model, with $\Omega_{m}\simeq0.27$, we get
Fig.~\ref{figura3}. The coincidence is very good and it is
difficult to distinguish between the two models.

\begin{figure}
[ptb]
\begin{center}
\includegraphics[
height=2.2001in, width=3.3797in
]%
{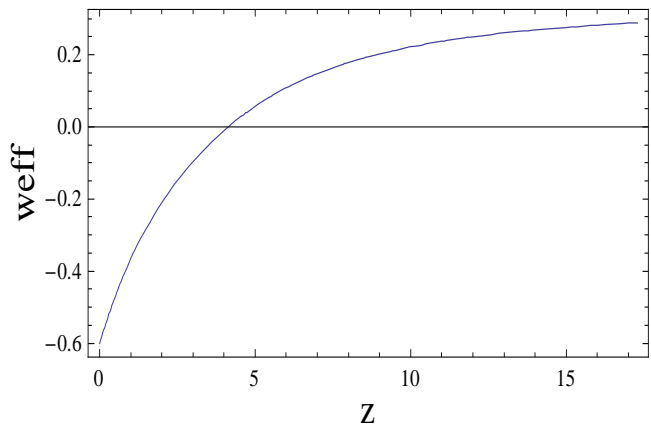}%
\caption{$w_{eff}$ versus redshift derived from the model.}\label{figura4}%
\end{center}
\end{figure}

\begin{figure}
[ptb]
\begin{center}
\includegraphics[
height=2.2001in, width=3.3797in
]%
{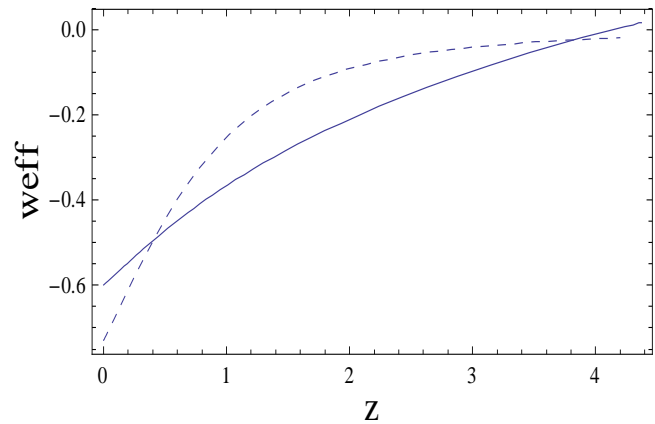}%
\caption{Comparison of the effective equation of state parameter
$w_{eff}$ for our model (continous) and $\Lambda CDM$ (dashed).
For $z$ larger than 4,
the radiation epoch should be carefully considered.}\label{figura5}%
\end{center}
\end{figure}

 Despite these good results,  some comments are in order. As we
have seen, in our model, the dynamical history of the universe is
described by the scale factor $a(t)\sim t^{1/2}$ at early epochs
and $a(t)\sim t^{2}$ at late times giving rise to a
matter-dust-like stage at intermediate times. This behavior
addresses, in principle, the two main issues of dark energy
models: $i)$ producing a Friedmann-like epoch suitable for LSS
formation and $ii)$ an accelerated present epoch stage. In Figs.
\ref{figura4} and ~\ref{figura5}, we have plotted the behavior of
the effective equation of state parameter
\begin{equation}
w_{eff}=-1-\frac{2}{3}\frac{\dot{H}}{H^2}\,,\end{equation} for our
model and compared it with the $\Lambda$CDM model. Clearly, also
if the model is accelerating at present epoch ($z\sim 0$), the
power is not enough to completely fit the prescription
$w_{eff}\simeq -1$ for the cosmological constant (see
Fig.~\ref{figura5}). However also  the $\Lambda$CDM model  does
not produce exactly $w_{eff}=1$ (since there is also the matter
component); in fact, if we consider $\Omega_{m}\simeq 0.27$, we
have $w_{eff}\simeq -0.73$ (in the case $\Omega_{m}=0.3$,
$w_{eff}=-0.7$). Therefore, the value of our $w_{eff}$ (absolute
value) is smaller than the desiderated value, but if we compare
this with $w_{eff}$ of the $\Lambda$CDM, it is not so far as if we
compare it with the pure $\Lambda$-case $w_{\Lambda}=-1$.

Furthermore, radiation should be included into dynamics. This fact
could destroy the nice feature achieved here, i.e. the smooth
transition between an unstable dust epoch to a  stable, asymptotic
accelerated phase. In this perspective, more accurate models,
including e.g. non-local gravitational corrections, should be
taken into account as done in \cite{Box}.

Finally, our discussion takes into account only the background
while fluctuations are not considered. In fact, at the background
level, we are able to obtain matter-like regime but things could
not work when fluctuations are included so  one should try to
mimic matter - like behavior by modifying gravity or including a
dynamical equation of state similar to the Chaplygin gas model
which well address this goal. This will be the argument of future
investigations. 

In summary, we have shown that suitable values of the parameters
in the presented general solution (\ref{solution}) allow to
reproduce the requested behavior of a Friedmann dust - like
solution evolving into an accelerated behavior as prescribed by
observations. This model, physically consistent, has been derived
by asking for a Noether symmetry in the $f(R)$ function. The
existence of such a symmetry fixes the form of $f(R)$ and allows
physically viable models.  However, starting from this approach,
more accurate models should be considered in order to address all
the issues related to the theory of perturbations and the
observational data sets.
%
%TCIMACRO{\FRAME{ftbpFU}{3.3797in}{2.2001in}{0pt}{\Qcb{Comparison of the
%distance modulus $\delta$. Of our model (continous) and $\Lambda CDMT$
%(dashed). The coincidence is almost perfect.}}{}{figura3.eps}%
%{\special{ language "Scientific Word";  type "GRAPHIC";
%maintain-aspect-ratio TRUE;  display "USEDEF";  valid_file "F";
%width 3.3797in;  height 2.2001in;  depth 0pt;  original-width 3.333in;
%original-height 2.1603in;  cropleft "0";  croptop "1";  cropright "1";
%cropbottom "0";  filename 'figura3.eps';file-properties "XNPEU";}}}%
%BeginExpansion
%EndExpansion

\section*{Acknowledgements}

PMM gratefully acknowledges the financial support provided by the I3P
framework of CSIC and the European Social Fund.


\newpage

\begin{thebibliography}{99}

\bibitem {LambdaTest}Seljak U. et al. 2005, Phys. Rev. D, 71, 103515

\bibitem {LambdaRev}Carroll S.M., Press W.H., Turner E.L. 1992, Ann. Rev.
Astron. Astroph., 30, 499

\bibitem {PR03}Peebles P.J.E., Rathra B. 2003, Rev. Mod. Phys., 75, 559;
Padmanabhan T. 2003, Phys. Rept., 380, 235

\bibitem {copeland}E.J. Copeland, M. Sami, S. Tsujikawa, \textit{Int. J. Mod.
Phys.} \textbf{D 15}, 1753 (2006).

\bibitem {will}C.~M.~Will,
%``The confrontation between general relativity and experiment,''
Living Rev. Relativity \textbf{9} (2006), arXiv:gr-qc/0510072.

\bibitem {kleinert}H. Kleinert, H.-J. Schmidt, \textit{Gen. Relativ.
Grav.}\ \textbf{34} 1295 (2002).

\bibitem {noi}Capozziello S. 2002, Int. J. Mod. Phys. D, 11, 483.\newline
Capozziello S., Carloni S., Troisi A. 2003, Rec. Res. Dev. in
Astron. and Astroph., 1, 1,
(arXiv\,:\,astro\,-\,ph/0303041).\newline Odintsov S.D., Nojiri S.
2003, Phys. Lett. B, 576, 5\newline Capozziello S., Cardone V.F.,
Carloni S., Troisi A. 2003, Int. J. Mod. Phys. D, 12,
1969.\newline Carroll S.M., Duvvuri V., Trodden M., Turner M.
2004, Phys. Rev. D, 70, 043528.\newline Allemandi G., Borowiec A.,
Francaviglia M. 2004, Phys. Rev. D, 70, 103503.\newline Nojiri S.
and Odintsov S.D. 2004, Gen. Rel. Grav. 36, 1765.\newline Cognola
G., Elizalde E., Nojiri S., S.D. Odintsov, Zerbini S. 2005, JCAP,
010.

\bibitem {odirev}S.Nojiri and S.D. Odintsov, \textit{Int. J. Meth. Mod. Phys.}
\textbf{4}, 115 (2007).

\bibitem {GRGrev}S. Capozziello and M. Francaviglia, \textit{Gen. Rel.
Grav.:} Dark Energy special issue, \textbf{40}, 357 (2008).


\bibitem {noipla}Capozziello S., Cardone V.F., Carloni S., Troisi A. 2004,
Phys. Lett. A, 326, 292

\bibitem {mond}Milgrom M. 1983, Astroph. Journ., 270, 365; \newline Bekenstein J. 2004,
Phys. Rev. D, 70, 083509

\bibitem {jcap}S. Capozziello, V.F. Cardone and A. Troisi JCAP \textbf{08},
001 (2006).

\bibitem {mnras}S. Capozziello, V.F. Cardone, A. Troisi,
Mon.\ Not.\ Roy.\ Astron.\ Soc.\ \textbf{375}, 1423 (2007).

\bibitem {sobouti}Y. Sobouti, A\&A, 464, 921 (2007).

\bibitem {salucci} C.Frigerio Martins
and P. Salucci, MNRAS 381, 1103, (2007).

\bibitem {mendoza}S. Mendoza and Y.M. Rosas-Guevara, A\&A, 472, 367 (2007).

\bibitem {instabilities-f(R)}V.~Faraoni, Phys.\ Rev.\ \textbf{D 72}, 124005
(2005); \newline G.~Cognola and S.~Zerbini, J.\ Phys.\ \textbf{A
39}, 6245 (2006); \newline G. Cognola, M. Gastaldi and S. Zerbini,
arXiv: gr\,-\,qc/0701138.

\bibitem {ghost-f(R)}K.~S.~Stelle, Gen.\ Rel.\ Grav.\ \textbf{9}, 353 (1978).

\bibitem {mimicking}Capozziello S., Cardone V.F., Troisi A. 2005, Phys. Rev.
D, 71, 043503

\bibitem {Hu}W.~Hu and I.~Sawicki,
%``Models of f(R) Cosmic Acceleration that Evade Solar-System Tests,''
Phys.\ Rev.\ D \textbf{76}, 064004 (2007).

\bibitem {Star}A.~A.~Starobinsky,
%``Disappearing cosmological constant in f(R) gravity,''
JETP Lett.\ \textbf{86}, 157 (2007).

\bibitem {Odintsov1}S. Nojiri and S.D. Odintsov, (2007) arXiv:0706.1378 [hep-th].

\bibitem {cimento} S. Capozziello, R. de Ritis, C. Rubano, and P. Scudellaro,
\textit{La Rivista del Nuovo Cimento} \textbf{4} (1996) 1.


\bibitem {leandros} S. Capozziello, S. Nesseris, L. Perivolaropoulos,
 JCAP {\bf 12}, 009 (2007).

\bibitem {lambiase} S. Capozziello and G. Lambiase, Gen.Rel. Grav. \textbf{32},
295 (2000).

\bibitem {CNOT} S. Capozziello, S. Nojiri, S.D. Odintsov, A. Troisi
\textit{Phys. Lett.}\ \textbf{B639}, 135 (2006).

\bibitem {Box} S. Jhingan, S. Nojiri, S.D. Odintsov, M. Sami, I.
Thongkool, and S. Zerbini, (2008) arXiv:0803.2613 [hep-th].




\end{thebibliography}
\end{document}